\documentclass[aps,prl,reprint,superscriptaddress,amsmath,amssymb,showpacs]{revtex4-1}
\usepackage{graphicx}
\usepackage{color}
\usepackage{bbm}

\date{\today}                  

\begin{document}

\title{Topological invariants to characterize universality of boundary charge in one-dimensional
insulators beyond symmetry constraints}

\author{Mikhail Pletyukhov}
\affiliation{Institut f\"ur Theorie der Statistischen Physik, RWTH Aachen, 
52056 Aachen, Germany and JARA - Fundamentals of Future Information Technology}
\author{Dante M. Kennes}
\affiliation{Institut f\"ur Theorie der Statistischen Physik, RWTH Aachen, 
52056 Aachen, Germany and JARA - Fundamentals of Future Information Technology}
\author{Jelena Klinovaja}
\affiliation{Department of Physics, University of Basel, Klingelbergstrasse 82, 
CH-4056 Basel, Switzerland}
\author{Daniel Loss}
\affiliation{Department of Physics, University of Basel, Klingelbergstrasse 82, 
CH-4056 Basel, Switzerland}
\author{Herbert Schoeller}
\email[Email: ]{schoeller@physik.rwth-aachen.de}
\affiliation{Institut f\"ur Theorie der Statistischen Physik, RWTH Aachen, 
52056 Aachen, Germany and JARA - Fundamentals of Future Information Technology}

\begin{abstract}

In the absence of any symmetry constraints we address universal properties of the 
boundary charge $Q_B$ for a wide class of nearest-neighbor tight-binding models 
in one dimension with one orbital per site but generic modulations of on-site potentials 
and hoppings. We provide a precise formulation of the bulk-boundary correspondence 
relating the boundary charge of a single band uniquely to the Zak phase evaluated in a 
particular gauge. We reveal the topological nature of $Q_B$ by
proving the quantization of a topological index $eI=\Delta Q_B - \bar{\rho}$, 
where $\Delta Q_B$ is the change of $Q_B$ when shifting the lattice by one site 
towards a boundary and $\bar{\rho}$ is the average 
charge per site. For a single band we find this index to be given by the winding
number of the fundamental phase difference of the Bloch wave function between the two
lattice sites defining the boundary of a half-infinite system. For a given chemical 
potential we establish a central topological constraint $I\in\{-1,0\}$ related only to 
charge conservation of particles and holes. Our results are shown to be stable against 
disorder and we propose generalizations to multi-channel and interacting systems.

\end{abstract}

\maketitle

{\it Introduction---}Motivated by the discovery of the Quantum Hall effect (QHE)
\cite{klitzing_dorda_pepper_prl_80,thouless_etal_prl_82}, the search for materials with 
topological edge states (TESs) has become a very important field of condensed matter 
physics and quantum optics 
[\onlinecite{volkov_pankratov_JETP_85}-\onlinecite{hsieh_etal_nature_08}], see 
Refs.~[\onlinecite{hasan_kane_RMP_10}-\onlinecite{asboth_book_16}] for reviews and textbooks. 
Routinely, topological insulators (TIs) are classified via their symmetry class and 
dimension [\onlinecite{schnyder_etal_prb_08}-\onlinecite{diez_etal_njp_15}].
Topological invariants like Chern and winding numbers are established and can be used to 
predict TESs at the boundary of two materials with different topological indices. Recently, 
the classification has been extended to include inversion symmetry within the field of 
topological crystalline insulators (TCIs) [\onlinecite{hughes_etal_prb_83}-\onlinecite{lau_etal_prb_16}]. 
Here, the Zak phase \cite{zak} is the topological invariant which, via the so-called 
modern theory of polarization (MTP), can be related to the boundary charge $Q_B$ 
[\onlinecite{kingsmith_vanderbilt_prb_93}-\onlinecite{vanderbilt_book_2018}], 
a relation called the surface charge theorem (SCT). 
However, since the Zak phase of an individual band is not gauge invariant an 
unknown integer of topological nature occurs in the SCT. Away from symmetry restrictions, finite 
one-dimensional (1D) tight-binding models with a sinusoidal on-site potential were studied 
\cite{park_etal_prb_16,thakurathi_etal_prb_18}, where a continuous phase variable $\varphi$ 
controls the offset of the potential. Surprisingly, in the long wavelength limit, $Q_B(\varphi)$ reveals a 
universal linear slope which was shown to be stable against disorder and to be related to the 
quantized Hall conductance. The linear behavior can be explained 
from classical charge conservation which, however, leaves again an unknown integer undetermined. 
Shifting the lattice adiabatically by {\it one site} towards a boundary of a 
half-infinite system, the boundary charge changes by the constant amount 
$\Delta Q_B=\bar{\rho}$ ($\text{mod}(e)$), where  $\bar{\rho}$ is the
average charge per site. This is a generalization of charge pumping discussed
extensively within the QHE \cite{thouless_prb_83,hatsugai_fukui_prb_16}, where the lattice is 
shifted by a {\it whole unit cell} such that the charge $\nu e$, given by the
number $\nu$ of occupied bands, is shifted into the boundary and balanced by a corresponding 
number of edge states leaving the band.

These works raise two important, fundamental issues that are intimately related: (i) the unknown 
integer in the boundary charge needs to be characterized and (ii) the topological nature of $Q_B$ 
and the relevance of symmetries should be addressed. This letter solves both of these issues
by first fixing the unknown integer in the SCT, a long outstanding problem in the MTP
(addressing (i)). This provides a precise formulation of the bulk-boundary 
correspondence relating $Q_B$ to bulk properties. Addressing (ii) within our framework 
we show that  $eI=\Delta Q_B-\bar{\rho}$, an invariant defined via shifting the lattice by 
{\it one site} towards a boundary, is quantized even beyond symmetry constraints. 
Our results motivate the boundary charge to be an interesting physical
observable bridging the research fields of TIs, the MTP, and the QHE.

We advance the description of $I$ in two central ways for a wide class of tight-binding 
models with non-degenerate bands in 1D. The first central result relates to $Q_B^{(\alpha)}$ 
of a \emph{single band} $\alpha$ and states that $I_\alpha=-w_\alpha\in\{0,\pm 1\}$, where the 
quantized and gauge invariant winding number $w_\alpha$ is defined in terms of the fundamental 
phase-difference of the Bloch wave function across two adjacent sites to the  right and 
left of the boundary defining a half-infinite system. We stress that this winding number 
can be accessed experimentally by measuring the charge via charge sensors and scanning
single-electron transistor microscopy techniques \cite{charge_sensors}. 
Furthermore, we show that $w_\alpha$ contains more information than the Chern number and 
is related to the Zak phase only in case of inversion symmetry. The second 
central result is obtained for the \emph{total invariant} $I$ for given chemical potential 
$\mu$ in some gap. Here, we show that charge conservation of particles and holes 
implies the topological constraint $I\in\{0,-1\}$ enforcing a corresponding constraint 
for the phase-dependence of the edge states. All of our results are demonstrated to be 
stable against random disorder (breaking translational invariance) and we propose 
generalizations to multi-channel and interacting systems. 

In an accompanying article \cite{paper_prb} we derive the central result $I\in\{-1,0\}$ 
rigorously by studying directly the conditions for the appearance of edge states from a 
convenient analytic continuation of Bloch states. The agreement with our physically 
motivated presentation in this letter reveals the surprising result that edge states are 
not the driving force for the constraint but have to adjust to a certain choice of 
the phase-dependence of the model parameters to respect charge conservation 
of particles and holes.

\begin{figure}
\centering
 \includegraphics[width=\columnwidth]{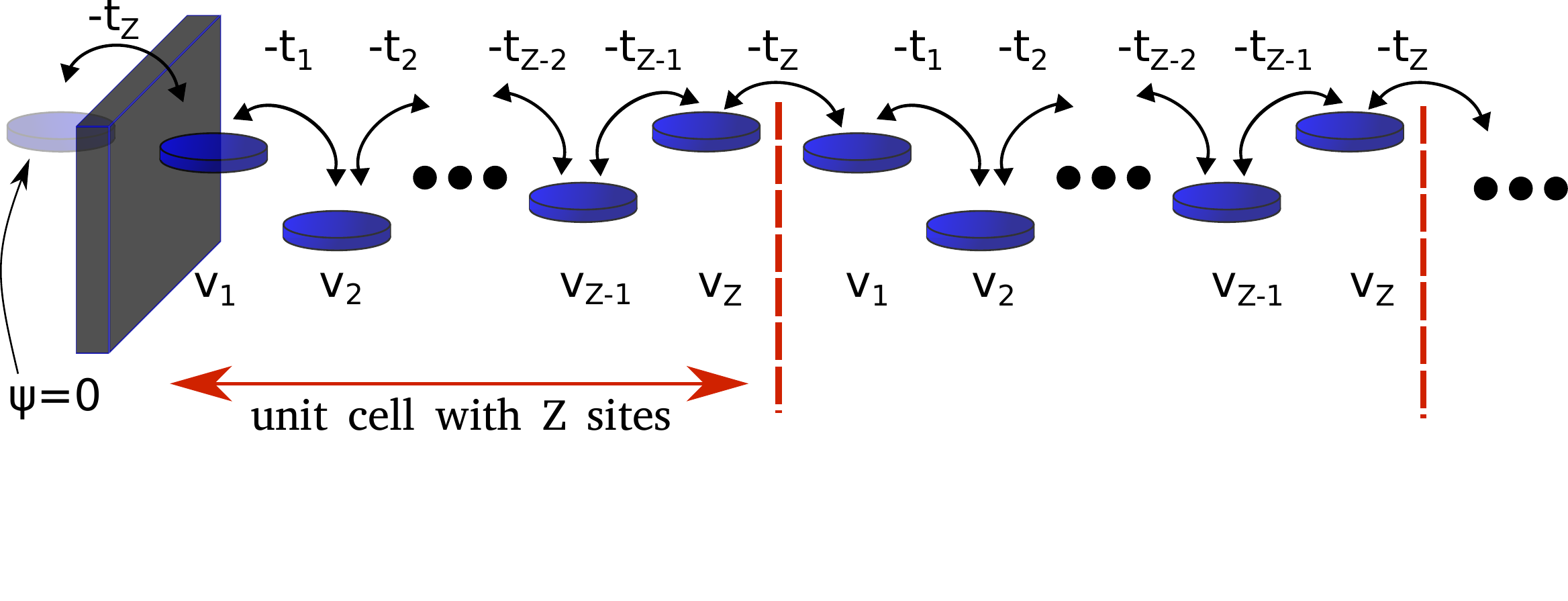}
 \caption{Sketch of the model. $Z$ denotes the number of sites within a unit cell. 
   The black bar indicates the left boundary such that the eigenstate of the infinite 
   system has to vanish on the site left to the boundary. The index $m=1,2,\dots$ labels
   the lattice sites starting from the boundary. 
 }
\label{fig:system}
\end{figure}

{\it Model---}
We start with a generic nearest-neighbor tight-binding model with translational invariance 
and one orbital per site on a half-infinite system, see Fig.~\ref{fig:system} for a sketch 
of the system (for more general setups, see Ref.~\onlinecite{paper_prb}). 
The Hamiltonian is given by $H=\sum_{m=1}^{\infty} \left\{v_m |m\rangle\langle m|
- (t_m |m+1\rangle\langle m| + h.c.)\right\}$, with generic on-site potentials
$v_m=v_{m+Z}$ and hoppings $t_m=t_{m+Z}$, where $Z$ is the number of sites within a 
unit cell. All $t_m$ are chosen real since the phases can be gauged away 
by a unitary transformation without changing the density \cite{paper_prb}. We consider zero temperature and use 
units $\hbar=e=a=1$, where $a$ denotes the lattice spacing between adjacent sites.

To address the central issue of how the properties of the system depend on the 
definition of the boundary we introduce a phase variable $0\le\varphi< 2\pi$ 
controlling the continuous shift of the lattice towards the boundary via 
$\gamma_{m+1}(\varphi)=\gamma_m(\varphi+{2\pi\over Z})$, with $\gamma=v,t$,
such that a phase change by ${2\pi\over Z}$ corresponds to a shift of the boundary by one site. 
Generically, we take the form $v_m(\varphi)=V F_v(\varphi+2\pi m/Z)$ and
$t_m(\varphi)=t+\delta t F_t(\varphi+2\pi m/Z)$, with 
$F_\gamma(\varphi)=F_\gamma(\varphi+2\pi)\sim O(1)$. For the special case of a single 
cosine modulation our model is equivalent to the well-known
generalized Aubry-Andr\'e-Harper models \cite{AAH} which play a central role in the 
study of TIs.

{\it Definition and decomposition of boundary charge---} 
In the insulating regime, the density $\rho(m)$ of a half-infinite system 
approaches the $Z$-periodic bulk value $\rho_{\text{bulk}}(m)$ 
of the infinite system exponentially fast \cite{paper_prb,kallin_halperin_prb_84} 
on a scale $\xi\sim t/\Delta$, where $t$ is the average hopping and $\Delta$ denotes 
the gap (motivating the choice of a half-infinite setup). Following 
Refs.~\cite{vanderbilt_book_2018,park_etal_prb_16}, we define the boundary charge 
as a macroscopic average 
$Q_B=\sum_{m=1}^\infty\left[\rho(m)-\bar{\rho}\right]f(m)$ of the excess density modeling 
a charge measurement probe characterized by some envelope function $f(m)$ decaying 
slowly from unity to zero compared to the scales $Z$ and $\xi$. Here, 
$\bar{\rho}={1\over Z}\sum_{j=1}^Z\rho_{\text{bulk}}(j)$ denotes the
average particle charge per site. Separating 
$\rho(m)-\bar{\rho}=\left[\rho(m)-\rho_{\text{bulk}}(m)\right]
+\left[(\rho_{\text{bulk}}(m)-\bar{\rho}\right]$, 
the first term of $Q_B$ leads to 
$\sum_{m=1}^\infty\left[\rho(m)-\rho_{\text{bulk}}(m)\right]$ since $f(m)\approx 1$
on the scale $m\lesssim\xi$. The second term gives
$Q_P=\sum_{n=0}^\infty\sum_{j=1}^Z\left[\rho_{\text{bulk}}(j)-\bar{\rho}\right]f(Zn+j)$, with
$f(Zn+j)\approx f(Zn)+ f'(Zn)j$. 
Using $Z\sum_{n=0}^\infty f'(Zn)\approx \int_0^\infty f'(x)=-1$
and $\sum_{j=1}^Z\left[\rho_{\text{bulk}}(j)-\bar{\rho}\right]=0$, we get  
$Q_P=-{1\over Z}\sum_{j=1}^Z j\left[\rho_{\text{bulk}}(j)-\bar{\rho}\right]$, describing
the negative bulk dipole moment per unit cell, in analogy to the 
surface polarization charge of a dielectric medium. Finally, separating
$\rho(m)=\rho_{\text{band}}(m)+\rho_{\text{edge}}(m)$ into the contribution of the bands and the
edge states, and defining the Friedel density via 
$\rho_F(m)=\rho_{\text{band}}(m)-\rho_{\text{bulk}}(m)$, we find the following gauge-invariant 
decomposition of the boundary charge 
\begin{align}
\label{eq:bbc}
Q_B = Q_F + Q_P + Q_E \,,
\end{align}
where $Q_F=\sum_{m=1}^\infty\rho_F(m)$ and $Q_E$ is the number of edge states. 
Interestingly, the form $Q_B-Q_E= Q_F + Q_P$ suggests a 
{\it bulk-boundary correspondence}, $Q_B-Q_E$ referring to the boundary, and 
$Q_F+Q_P$ containing the bulk properties. 
\begin{figure}
\centering
 \includegraphics[width=\columnwidth]{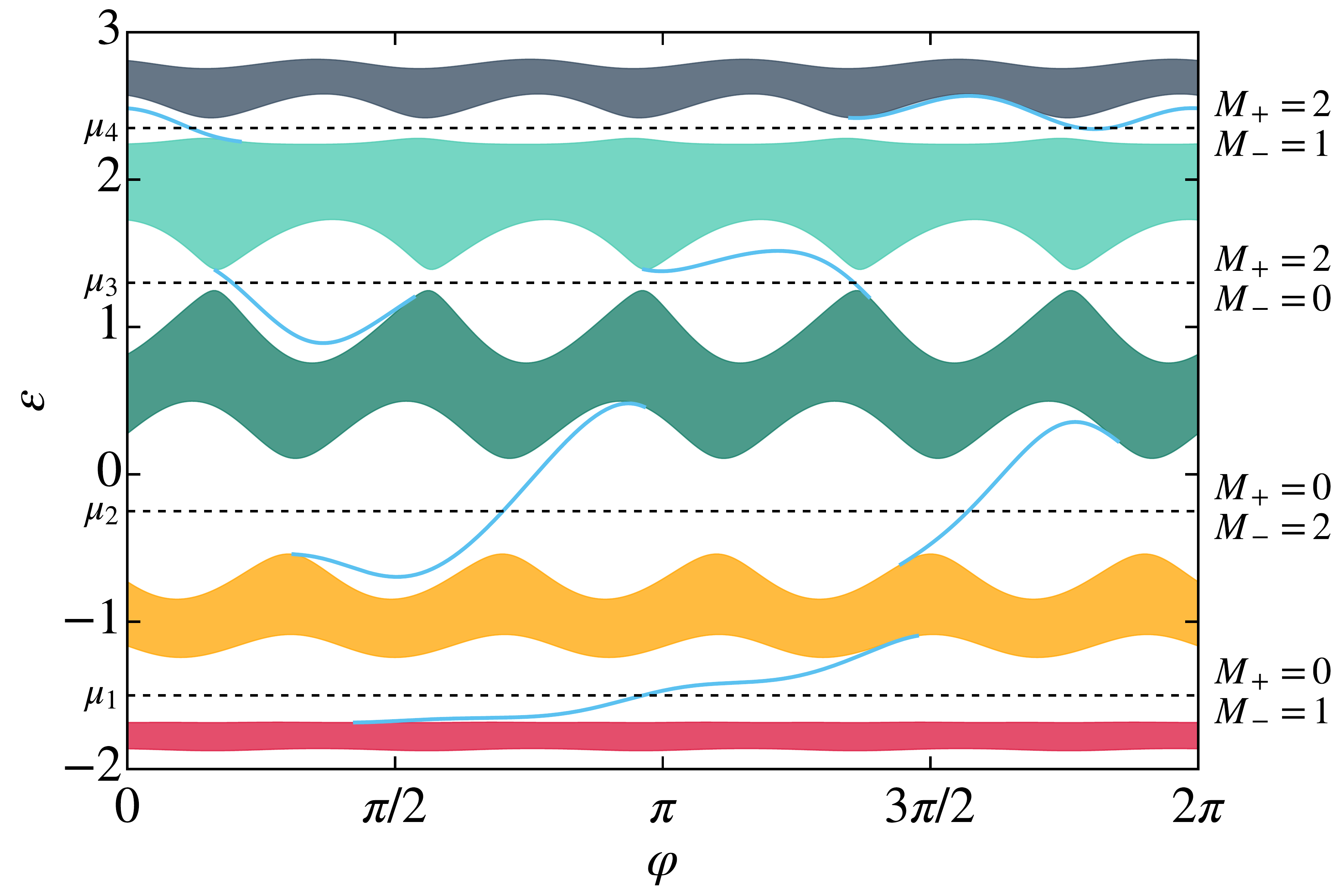}
 \caption{Illustration of the band structure and of the phase-dependence of the 
   edge states (blue color) 
   connecting the bands for $Z=5$, $t=1$, $V=0.5$, $\delta t=0.1$, and 
   $F_\gamma(\varphi)$ defined via three random Fourier components for $\gamma=v,t$, 
   see Supplemental Material \cite{SM} for the concrete parameters. 
   The chemical potentials $\mu_\nu$ in  
   gap $\nu$ are indicated by dashed horizontal lines, for which $Q_B$ is 
   calculated in Fig.~\ref{fig:QB_invariant}.
   To the right we state the total numbers $M_\pm(\mu_\nu)$ of edge states entering/leaving 
   the system corresponding to the four chemical potentials $\mu_\nu$.
}
\label{fig:band_structure}
\end{figure}
\begin{figure}
\centering
 \includegraphics[width=\columnwidth]{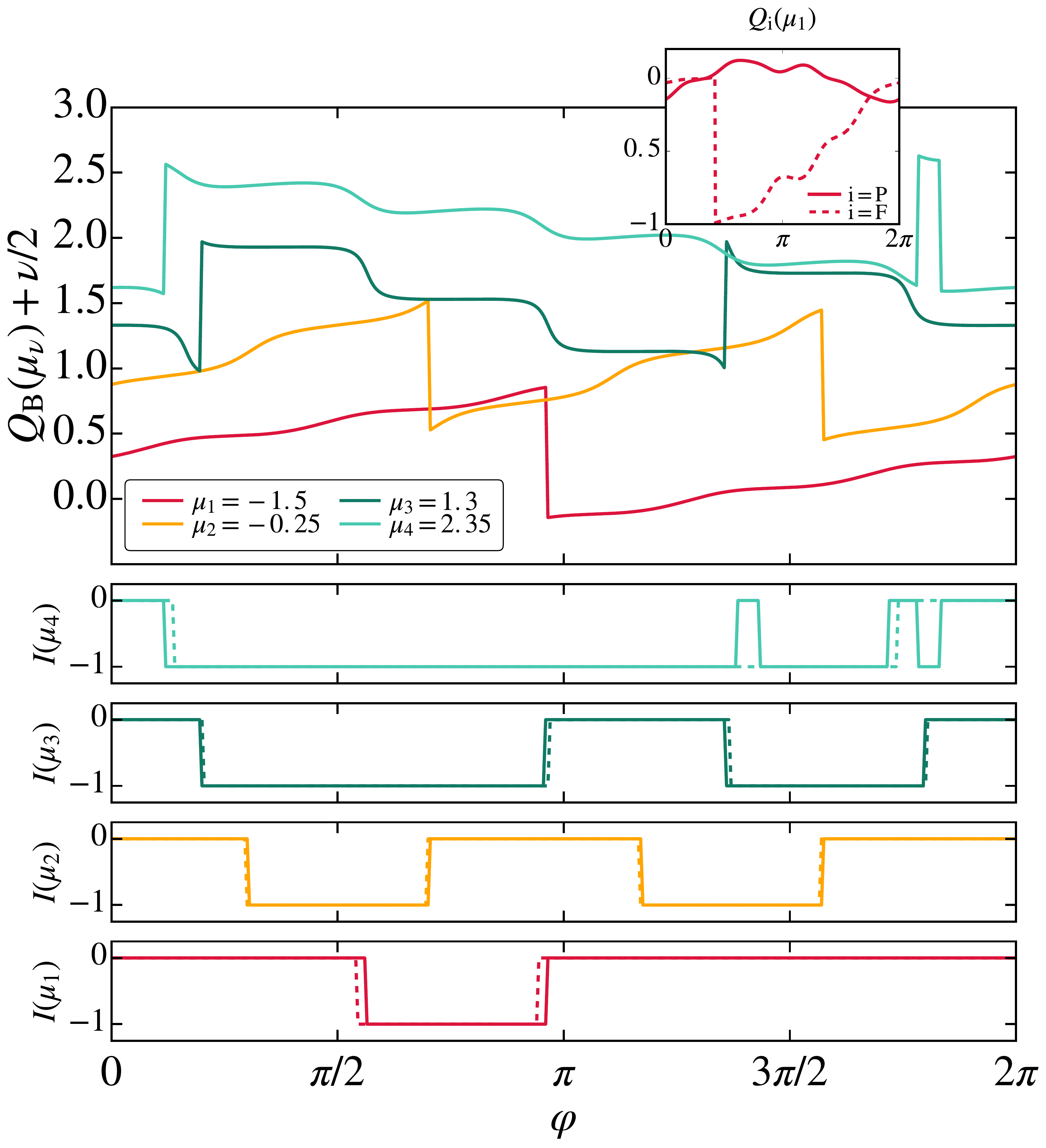}
 \caption{Boundary charge $Q_B$ and the invariant $I$ as function of $\varphi$ for a 
   half-infinite system using the parameters of Fig.~\ref{fig:band_structure} for 
   several $\mu_\nu$. 
   We show $Q_B+\nu/2$ to offset the different curves. Up to a $2\pi/Z$-periodic function, 
   $Q_B$ shows on average a linear slope with jumps at the positions where edge states 
   move above/below $\mu_\nu$. As shown in the right inset, $Q_F$ or $Q_P$
   alone do not show any linear behaviour. The invariant is always quantized to 
   $I\in\{-1,0\}$. The dashed line is the invariant with additional staggered
   onsite-disorder drawn from a uniform distribution $(0,0.05]$ for a very large 
   finite system of $5\cdot 10^5$ lattice sites. In the lowest panel we show the invariant
   $I_\alpha$ with $\alpha=2$ for a single band. It can only take the values 
   $I_\alpha\in\{0,\pm 1\}$.}
\label{fig:QB_invariant}
\end{figure}
{\it Surface charge theorem---} We now express $Q_F$ and $Q_P$ via the Bloch 
states of the infinite system. The bulk spectrum consists of $Z$ non-degenerate 
bands $\alpha=1,\dots,Z$ (numerated from bottom to top) with $Z-1$ gaps 
$\nu=1,\dots,Z-1$ in between. We assume that each gap remains open for all $\varphi$;
see Fig.~\ref{fig:band_structure} for an illustration of the $\varphi$-dependence of 
the band structure. If the chemical potential $\mu=\mu_\nu$ is somewhere in gap $\nu$, 
we get $\bar{\rho}=\nu/Z$ and can split $Q_F=\sum_{\alpha=1}^\nu Q_F^{(\alpha)}$ and 
$Q_P=\sum_{\alpha=1}^\nu Q_P^{(\alpha)}$ into the contributions
of the occupied bands, where $Q_F^{(\alpha)}=\sum_{m=1}^\infty\rho_F^{(\alpha)}(m)$ and 
$Q_P^{(\alpha)}=-{1\over Z}\sum_{j=1}^Z j(\rho_{\text{bulk}}^{(\alpha)}(j)-{1\over Z})$. 
The densities $\rho^{(\alpha)}_\text{band}(m)=\int_0^{\pi/Z}dk|\psi_k^{(\alpha)}(m)|^2$ and 
$\rho^{(\alpha)}_\text{bulk}(m)=\int_{-\pi/Z}^{\pi/Z}dk|\psi_{k,\text{bulk}}^{(\alpha)}(m)|^2$
can be expressed by the eigenstates $\psi_k^{(\alpha)}$ and $\psi_{k,\text{bulk}}^{(\alpha)}$ of 
the half-infinite and infinite system, respectively, where $k$ denotes the quasi-momentum. 
Using the Bloch form 
$\psi^{(\alpha)}_{k,\text{bulk}}(m) = \left({Z \over 2\pi}\right)^{1\over 2}u_k^{(\alpha)}(m)e^{ikm}$,
we find $\psi_k^{(\alpha)}(m)=\psi^{(\alpha)}_{k,\text{bulk}}(m)- {\rm c.c.}$, 
where $u_k^{(\alpha)}(m)= u_k^{(\alpha)}(m+Z)$
are the periodic Bloch states (normalized to a unit cell) with 
$u_k^{(\alpha)}(Z)=\left({Z\over 2\pi}\right)^{1\over 2}\psi_{k,\text{bulk}}^{(\alpha)}(0)$ chosen real in order to 
fulfill the boundary condition $\psi_k^{(\alpha)}(0)=0$. Together with the periodic gauge condition
$\psi^{(\alpha)}_{k,\text{bulk}}=\psi^{(\alpha)}_{k+{2\pi\over Z},\text{bulk}}$
this fixes uniquely the gauge of the Bloch states. Using the eigenstates we finally 
express $\rho_F^{(\alpha)}(m)=-{Z\over 2\pi}\int_{-\pi/Z}^{\pi/Z} dk [u_k^{(\alpha)}(m)]^2 e^{2ikm}$ 
and $\rho_{\text{bulk}}^{(\alpha)}(m)={Z\over 2\pi}\int_{-\pi/Z}^{\pi/Z} dk |u_k^{(\alpha)}(m)|^2$ 
via the Bloch states, providing the charges $Q^{(\alpha)}_F$ and $Q^{(\alpha)}_P$. 
Using recursion relations for $u_k^{(\alpha)}(m)$, provided by the nearest-neighbor
hopping Hamiltonian, we show in Ref.~\cite{paper_prb} that 
$Q_B^{(\alpha)}=Q^{(\alpha)}_F+Q^{(\alpha)}_P=-{\bar{\gamma}_\alpha\over 2\pi} + P_{\text{ion}}$ 
can be expressed via the Zak phase $\bar{\gamma}_\alpha= i\int_{-\pi/Z}^{\pi/Z} dk 
\sum_{j=1}^Z[u^{(\alpha)}_k(j)]^*{d\over d k} u^{(\alpha)}_k(j)$
introduced in Ref.~\cite{zak}, and the ionic part of the polarization
$P_\text{ion}={1\over Z^2}\sum_{j=1}^Z(j-Z)$. Remarkably, this relation holds only exactly when the 
gauge of $u^{(\alpha)}_k$ is chosen such that 
$u^{(\alpha)}_k(Z)$ is real which is 
fundamentally related to the boundary condition. This result provides a {\it unique} 
formulation of the SCT relating the Zak phase to the boundary charge of a single band. 
We emphasize that the proper gauge is an essential 
ingredient to fix the unknown integer of $Q_B-Q_E=\sum_{\alpha=1}^\nu Q_B^{(\alpha)}$ 
and to give Eq.~(\ref{eq:bbc}) a precise meaning in terms of the bulk-boundary correspondence.

{\it Invariant of a single band---}We now address the central issue how the boundary charge changes when we change the
phase by ${2\pi\over Z}$. We first analyze the boundary charge $Q_B^{(\alpha)}$ of a single 
band $\alpha$. The change $\Delta Q_B^{(\alpha)}(\varphi)=
Q_B^{(\alpha)}(\varphi+{2\pi\over Z})-Q_B^{(\alpha)}(\varphi)
=-{\Delta\bar{\gamma}_\alpha(\varphi)\over 2\pi}$ 
can be calculated from the corresponding change $\Delta\bar{\gamma}_\alpha$ of the
Zak phase which is related to the Bloch states $\tilde{u}_k^{(\alpha)}$ 
at phase $\varphi+{2\pi\over Z}$ of the shifted system. 
The latter follow from the shifted Bloch wave function 
$\tilde{\psi}^{(\alpha)}_{k,\text{bulk}}(m)=e^{-i\theta^{(\alpha)}_k}\psi^{(\alpha)}_{k,\text{bulk}}(m+1)$. 
Here, $\theta^{(\alpha)}_k$ denotes the phase of $\psi^{(\alpha)}_{k,\text{bulk}}(1)$ such 
that $\tilde{\psi}^{(\alpha)}_{k,\text{bulk}}(0)$ is real consistent with the boundary condition.
This gives $\tilde{u}_k^{(\alpha)}(m)=e^{-i\theta^{(\alpha)}_k}e^{ik}u_k^{(\alpha)}(m+1)$ 
leading to $\Delta\bar{\gamma}_\alpha=2\pi(w_\alpha-{1\over Z})$, where
$w_\alpha={1\over 2\pi i}\int_{-\pi/Z}^{\pi/Z} dk \,e^{-i\theta^{(\alpha)}_k}{d\over d k} e^{i\theta^{(\alpha)}_k}$
is the winding number of the phase factor $e^{i\theta^{(\alpha)}_k}$. We note that this winding 
number is integer and gauge invariant since $\theta^{(\alpha)}_k$ is the phase difference
of $\psi^{(\alpha)}_{k,\text{bulk}}(m)$ between the sites $m=1$ and $m=0$ defining the boundary.
As a result we can define an integer and universal invariant $I_\alpha$ for band $\alpha$
\begin{align}
\label{eq:invariant_alpha}
I_\alpha(\varphi)\equiv\Delta Q_B^{(\alpha)}(\varphi)-{1\over Z} = - w_\alpha(\varphi)\,.
\end{align}
As shown below $I_\alpha\in\{0,\pm1\}$ can only take three possible values, see 
Fig.~\ref{fig:QB_invariant}. 
Due to charge conservation $Q^{(\alpha)}_B$ will jump by $\pm 1$ when an edge state 
enters/leaves the band at $\varphi^{(\alpha)}_{i\pm}$. Therefore, $I_\alpha$ will jump by $\mp 1$ ($\pm 1$) at 
$\varphi=\varphi^{(\alpha)}_{i\pm}$ ($\varphi=\varphi^{(\alpha)}_{i\pm}-{2\pi\over Z}$).  
We conclude that $w_\alpha(\varphi)$ characterizes the {\it value} and the {\it jumps} 
of $I_\alpha(\varphi)$ in the whole phase interval, whereas the Chern number $C^{(\alpha)}$,
which is known to be the number of leaving minus the number of entering
edge states \cite{thouless_etal_prl_82,dana_jpc_85,kohmoto_prb_89_jpsj_92,hatsugai_prb_93}, 
is a measure for the {\it sum over all jumps} of $I_\alpha$
at the entering/leaving points $\varphi^{(\alpha)}_{i\pm}$ of the edge states. Therefore, 
$w_\alpha$ contains much more information than $C^{(\alpha)}$ and characterizes a different
physical quantity. In the special case of inversion symmetry 
($v_m=v_{Z-m+1}$ and $t_m=t_{Z-m}$) it can be shown that 
$\bar{\gamma}_\alpha=-\pi w_\alpha+2\pi P_{\text{ion}}$ \cite{paper_prb}. 
Therefore, $w_\alpha$ is the appropriate winding 
number to generalize the concepts of TCIs to cases without inversion symmetry.

{\it Total invariant---}Next we discuss the change of the total boundary charge $Q_B$ given by 
$\Delta Q_B=\sum_{\alpha=1}^\nu
\Delta Q_B^{(\alpha)}+\Delta Q_E=\sum_{\alpha=1}^\nu I_\alpha +{\nu\over Z}+\Delta Q_E$, 
where $\Delta Q_E$ is the change of the number of occupied edge states. This yields 
the result that the total invariant
\begin{align}
\label{eq:invariant_def} 
I(\varphi,\mu_\nu) \equiv \Delta Q_B(\varphi,\mu_\nu) - {\nu\over Z}
\end{align} 
is an integer {\it irrespective of any symmetry conditions}. As we have seen during 
the derivation the polarization charge $Q_P$
plays a very important role for this result. Only for chiral symmetry (all $v_m=0$) 
and half-filling or in case of inversion symmetry we get $Q_P=0$.

{\it Particle-hole duality---}We now present intuitive arguments why the invariant is integer and which values are
possible. Using charge conservation on average the particle charge $\bar{\rho}$ will 
be moved into the boundary when the system is shifted by one site, leading to 
$\Delta Q_B=\bar{\rho}$. Using the Pauli principle we find from charge conservation 
for the holes that on average the hole charge $\bar{\rho}_h=\bar{\rho}-1$ is moved 
into the boundary. Since the hole density is defined by 
$\rho_h(m)=\rho(m)-1$ the boundary charges for holes and particles are the same. Therefore, 
we obtain another value $\Delta Q_B=\bar{\rho}-1$ and conclude
\begin{align}
\label{eq:invariant_values}
\Delta Q_B(\varphi,\mu_\nu) \in \{\bar{\rho},\bar{\rho}-1\} \,\Leftrightarrow\,
I(\varphi,\mu_\nu) \in \{0,-1\} \,.
\end{align}
This provides also the result $I_\alpha\in\{0,\pm 1\}$ since $I_\alpha(\varphi)$ is the 
difference of $I(\varphi,\mu_\nu)$ when $\mu_\nu$ is chosen as the top or bottom of band $\alpha$. 
Which value occurs for a given phase depends crucially on the model parameters and can not
be predicted in general. To derive this result we have disregarded that during the shift edge 
states can cross $\mu_\nu$ so that 
(\ref{eq:invariant_values}) could in principle hold only $\text{mod}(1)$. However, the occurrence 
of edge states during the shift is rather artificial and depends crucially on the choice of the
phase-dependence on the interval $[\varphi,\varphi+{2\pi\over Z}]$ via $F_\gamma(\varphi)$. 
Therefore, the edge states are {\it not} the physical reason why $Q_B$ can only
change according to Eq.~(\ref{eq:invariant_values}). 
In Ref.~\cite{paper_prb} we show that, for given model parameters 
$v_m, t_m$ at phase $\varphi$ (and, correspondingly, at all phases shifted by 
${2\pi\over Z}$), the phase dependence can always be chosen in such a way 
that no edge state crosses $\mu_\nu$ in the phase interval $[\varphi,\varphi+{2\pi\over Z}]$. 
Therefore, since $\Delta Q_B(\varphi,\mu_\nu)$ depends only on the model parameters at $\varphi$,
we obtain the two values stated in Eq.~(\ref{eq:invariant_values}).

{\it Phase dependence of boundary charge---} 
We now determine the universal form of $Q_B(\varphi)$. Charge conservation dictates that 
$Q_B$ can only jump by $\pm 1$ at the phase points $\varphi_{i\pm}(\mu_\nu)$, where edge states 
move below/above $\mu_\nu$. Eq.~(\ref{eq:invariant_values}) implies that, up to a 
$2\pi/Z$-periodic and continuous function $f(\varphi,\mu_\nu)=f(\varphi+{2\pi\over Z},\mu_\nu)$, 
$Q_B$ is a linear function in $\varphi$ between the jumps. Therefore, we obtain the 
universal form (in agreement with Refs.~\cite{park_etal_prb_16,thakurathi_etal_prb_18})
\begin{align}
\label{eq:QB_form}
Q_B(\varphi,\mu_\nu) = f(\varphi,\mu_\nu)+{M_\nu\over 2\pi}\varphi + F(\varphi,\mu_\nu)\,,
\end{align}
with $F(\varphi,\mu_\nu)=\sum_{\sigma=\pm}\sum_{i=1}^{M_\sigma(\mu_\nu)}
\sigma\theta[\varphi-\varphi_{i\sigma}(\mu_\nu)]$, 
where $M_\pm(\mu_\nu)$ is the total number of edge states moving below/above $\mu_\nu$ 
when the phase changes by $2\pi$. The slope ${M_\nu\over 2\pi}$ of the linear term, with 
$M_\nu=M_-(\mu_\nu)-M_+(\mu_\nu)$, follows from the condition 
$Q_B(\varphi+2\pi,\mu_\nu)=Q_B(\varphi,\mu_\nu)$ and depends only on $\nu$ but not on
the precise position of $\mu_\nu$. We note that the unknown function $f(\varphi,\mu_\nu)$ 
in Eq.~(\ref{eq:QB_form}) is non-universal. Numerically, we observe that $f(\varphi,\mu_\nu)$ 
decreases for increasing $Z$ and is rather small for a smooth phase dependence of 
$v_m(\varphi), t_m(\varphi)$ and not too small gaps. Importantly, 
the phase-dependence of the edge states have to follow a topological constraint upon 
varying $\varphi$ such that $I\in\{0,-1\}$ is fulfilled. From 
Eqs.~(\ref{eq:invariant_values}) and (\ref{eq:QB_form}), we get
$M_\nu = \nu - s_\nu Z$, where $s_\nu=\Delta F(\varphi,\mu_\nu)-I(\varphi,\mu_\nu)$ 
depends only on $\nu$. Whereas $M_\nu$ is known to be
identical to the Chern number of the $\nu$ lowest bands 
\cite{thouless_etal_prl_82,dana_jpc_85,kohmoto_prb_89_jpsj_92,hatsugai_prb_93},
we find here, from the form of $s_\nu$ and $I\in\{0,-1\}$, the topological constraint
$\Delta F(\varphi,\mu_\nu)\in\{s_\nu-1,s_\nu\}$. This provides certain restrictions how the
edge states emerge as function of $\varphi$ which they have to follow in order to fulfil 
charge conservation of particles and holes compared to the case when the phase dependence is 
chosen such that no edge state crosses $\mu_\nu$ during the shift of the lattice by one site.

For the parameters of Fig.~\ref{fig:band_structure}, we 
show in Fig.~\ref{fig:QB_invariant} $Q_B(\varphi,\mu_\nu)$ 
and $I(\varphi,\mu_\nu)$ for various $\mu_\nu$, supporting our results stated in 
Eqs.~(\ref{eq:invariant_values}) and (\ref{eq:QB_form}). Importantly, it is essential
to consider the sum of $Q_F+Q_P$ for $Q_B$, neither $Q_F$ nor $Q_P$ alone 
lead to a linear slope, see the inset of Fig.~\ref{fig:QB_invariant}. Only for weak (strong) 
potentials $v_m\ll t_m$ ($v_m\gg t_m$) one finds that $Q_F$ ($Q_P$) dominates.
In Fig.~\ref{fig:QB_invariant} we also demonstrate the stability of our results 
against random disorder which breaks translational invariance. Most remarkably, due 
to the intuitive interpretation of our central result that 
$\Delta Q_B\in\{\bar{\rho},\bar{\rho}-1\}$ this is expected and holds also in the 
presence of weak interactions.

{\it Outlook---}Our results are 
derived for generic 1-channel tight-binding models, but we expect similar universal features
for multi-channel systems. Taking $N_c$ weakly coupled copies of the same system 
leads to the quantized values $I=\Delta Q_B-\bar{\rho}\in\{-N_c,\dots,-1,0\}$ of the invariant 
or $\Delta Q_B \in\{\bar{\rho}-N_c,\dots,\bar{\rho}-1,\bar{\rho}\}$,
with $\bar{\rho}=N_c\nu/Z$ \cite{multi-channel}. 
Our intuitive interpretation
for the occurrence of the allowed values for $\Delta Q_B$ suggest the results to be also stable
against weak interactions \cite{lin_etal_preprint} similar to the stability of bound states 
\cite{gangadharaiah_etal_prl_12}. 
Furthermore, via dimensional reduction, we expect our results to be also relevant to the 
understanding of universal features of $Q_B$ in higher dimensions. 
In the special case of half-filling gap closings can occur such that the average slope of the
linear term $M_\nu=0$. In this case Weyl physics occurs with interesting quantization of
the boundary charge itself \cite{weyl}.

We thank P. W. Brouwer, C. Bruder, F. Hassler, V. Meden, M. Thakurathi and S. Wessel for 
fruitful discussions. This work was supported by the Deutsche Forschungsgemeinschaft 
via RTG 1995, the Swiss National Science Foundation (SNSF) and NCCR QSIT. Simulations 
were performed with computing resources granted by RWTH Aachen University under project 
prep0010. Funding was received from the European Union's Horizon 2020 research, innovation 
program (ERC Starting Grant, grant agreement No 757725) as well as from the independence grant 
from the CRC 183 network.

M.P. and D.M.K. contributed equally to this work.


\begin{thebibliography}{99}


\bibitem{klitzing_dorda_pepper_prl_80}
K. v. Klitzing, G. Dorda, and M. Pepper, Phys. Rev. Lett. {\bf 45}, 494 (1980).

\bibitem{thouless_etal_prl_82}
D.J. Thouless, M. Kohmoto, M.P. Nightingale, and M. den Nijs, 
Phys. Rev. Lett. {\bf 49}, 405 (1982).



\bibitem{volkov_pankratov_JETP_85}
B.A. Volkov and O.A. Pankratov, Pis’ma Zh. Eksp. Teor. Fiz. {\bf 42}, 145
(1985) [JETP Lett. {\bf 42}, 178 (1985)].

\bibitem{pankratov_etal_SSC_87}
O.A. Pankratov, S.V. Pakhomov, and B.A. Volkov, Solid State Commun. {\bf 61}, 93 (1987).

\bibitem{kane_mele_prl_95}
C.L. Kane and E.J. Mele, Phys. Rev. Lett. {\bf 95}, 146802 (2005).

\bibitem{bernevig_etal_science_06}
B.A. Bernevig, T.L. Hughes, and S.-C. Zhang, Science {\bf 314}, 1757 (2006).

\bibitem{fu_kane_mele_prl_07}
L. Fu, C.L. Kane, and E.J. Mele, Phys. Rev. Lett. {\bf 98}, 106803 (2007).



\bibitem{koenig_etal_science_07}
M. K\"onig, S. Wiedmann, C. Brune, A. Roth, H. Buhmann, L.W. Molenkamp,
X.-L. Qi, and S.-C. Zhang, Science {\bf 318}, 766 (2007).

\bibitem{hsieh_etal_nature_08}
D. Hsieh, D. Qian, L. Wray, Y. Xia, Y. S. Hor, R. J. Cava, and M. Z. Hasan, 
Nature (London) {\bf 452}, 970 (2008).




\bibitem{hasan_kane_RMP_10}
M. Z. Hasan and C. L. Kane, Rev. Mod. Phys. {\bf 82}, 3045 (2010).

\bibitem{qi_zhang_RMP_11}
X.-L. Qi and S.-C. Zhang, Rev. Mod. Phys. {\bf 83}, 1057 (2011).

\bibitem{bernevig_book_13}
B.A. Bernevig, {\it Topological Insulators and Topological Superconductors}, 
Princeton University Press (2013).

\bibitem{tkachov_book_15}
G. Tkachov, {\it Topological Insulators: The Physics of Spin Helicity in Quantum Transport},
(Pan Stanford, 2015).

\bibitem{asboth_book_16}
J.K. Asb\'oth, L. Oroszl\'any, and A. P\'alyi, {\it A Short Course on Topological Insulators},
Lecture Notes in Physics, Springer 2016. 



\bibitem{schnyder_etal_prb_08}
A.P. Schnyder, S. Ryu, A. Furusaki, and A.W.W. Ludwig,
Phys. Rev. B {\bf 78}, 195125 (2008).

\bibitem{schnyder_etal_njp_10}
S. Ryu, A.P. Schnyder, A. Furusaki, and A.W.W. Ludwig,
New J. Phys. {\bf 12}, 065010 (2010).

\bibitem{kitaev_advphys_09}
A. Kitaev, in Advances in Theoretical Physics, edited by V. Lebedev and M. Feigel’man, 
AIP Conf. Proc. No. 1134 (AIP, New York, 2009), p. 22.

\bibitem{slager_etal_natphys_12}
R.-J. Slager, A. Mesaros, V. Juri\v{c}i\'c, and J. Zaanen, Nat. Phys. {\bf 9}, 98 (2012).

\bibitem{jadaun_etal_prb_13}
P. Jadaun, D. Xiao, Q. Niu, and S. K. Banerjee, Phys. Rev. B {\bf 88}, 085110 (2013).

\bibitem{chiu_etal_prb_13}
C.-K. Chiu, H. Yao, and S. Ryu, Phys. Rev. B {\bf 88}, 075142 (2013).

\bibitem{zhang_kane_mele_prl_13}
F. Zhang, C. L. Kane, and E. J. Mele, Phys. Rev. Lett. {\bf 111}, 056403 (2013).

\bibitem{benalcazar_etal_prb_14}
W.A. Benalcazar, J.C.Y. Teo, and T.L. Hughes, Phys. Rev. B {\bf 89}, 224503 (2014).

\bibitem{morimoto_furusaki_prb_13}
T. Morimoto and A. Furusaki, Phys. Rev. B {\bf 88}, 125129 (2013).

\bibitem{diez_etal_njp_15}
M. Diez, D.I. Pikulin, I.C. Fulga, and J. Tworzydlo, New J. Phys. {\bf 17}, 043014 (2015).



\bibitem{hughes_etal_prb_83}
T. L. Hughes, E. Prodan, and B. A. Bernevig,
Phys. Rev. B {\bf 83}, 245132 (2011).

\bibitem{chiu_etal_prb_88}
C.-K. Chiu, H. Yao, and S. Ryu, Phys. Rev. B {\bf 88}, 075142 (2013).

\bibitem{shiozaki_sato_prb_90}
K. Shiozaki and M. Sato, Phys. Rev. B {\bf 90}, 165114 (2014).

\bibitem{alexandradinata_etal_prb_16}
A. Alexandradinata, Zhijun Wang, and B.A. Bernevig, Phys. Rev. X {\bf 6}, 021008 (2016).

\bibitem{trifunovic_brouwer_prb_17}
L. Trifunovic and P. Brouwer, Phys. Rev. B {\bf 96}, 195109 (2017). 

\bibitem{lau_etal_prb_16}
A. Lau, C. Ortix, and J. van den Brink, Phys. Rev. Lett. {\bf 115}, 216805 (2015);
A. Lau, J. van den Brink, and C. Ortix, Phys. Rev. B {\bf 94}, 165164 (2016);
A. Lau and C. Ortix, Eur. Phys. J. Spec. Top. 227, 1309 (2018).



\bibitem{zak}
J. Zak, Phys. Rev. Lett. {\bf 48}, 359 (1982);
J. Zak, Phys. Rev. Lett. {\bf 62}, 2747 (1989).

\bibitem{kingsmith_vanderbilt_prb_93}
R.D. King-Smith and D. Vanderbilt, Phys. Rev. B(R) {\bf 47}, 1651 (1993);
D. Vanderbilt and R.D. King-Smith, Phys. Rev. B {\bf 48}, 4442 (1993).

\bibitem{resta_revmodphys_94}
R. Resta, Rev. Mod. Phys. {\bf 66}, 899 (1994).

\bibitem{kudin_etal_chemphys_07}
K.N. Kudin and R. Car, J. of Chem. Phys. {\bf 126}, 234101 (2007).

\bibitem{marzari_etal_revmodphys_12}
N. Marzari, A.A. Mostofi, J.R. Yates, I. Souza, and D. Vanderbilt, 
Rev. Mod. Phys. {\bf 84}, 1419 (2012).

\bibitem{spaldin_solidstatechem_12}
N.A. Spaldin, J. of Solid State Chem. {\bf 195}, 2 (2012).

\bibitem{rhim_etal_prb_17}
J.-W. Rhim, J. Behrends and J.H. Bardarson, Phys. Rev. B {\bf 95}, 035421 (2017).

\bibitem{miert_ortix_prb_17}
G. van Miert and Carmine Ortix, Phys. Rev. B {\bf 96}, 235130 (2017).

\bibitem{vanderbilt_book_2018}
D. Vanderbilt, {\it Berry Phases in Electronic Structure Theory: 
Electric Polarization, Orbital Magnetization and Topological Insulators},
(Cambridge University Press, 2018).












\bibitem{park_etal_prb_16}
J.-H. Park, G. Yang, J. Klinovaja, P. Stano, and D. Loss, Phys. Rev. B {\bf 94}, 075416 (2016). 

\bibitem{thakurathi_etal_prb_18}
M. Thakurathi, J. Klinovaja, and D. Loss, Phys. Rev. B {\bf 98}, 245404 (2018).




 \bibitem{thouless_prb_83}
 D.J. Thouless, Phys. Rev. B {\bf 27}, 6083 (1982).

 \bibitem{hatsugai_fukui_prb_16}
 Y. Hatsugai and T. Fukui, Phys. Rev. B {\bf 94}, 041102(R) (2016).



\bibitem{charge_sensors}
M.J. Yoo, T.A. Fulton, H.F. Hess, R.L. Willett, L.N. Dunkleberger, R.J. Chichester, 
L.N. Pfeiffer, and K.W. West, Science {\bf 276} 579 (1997);
S.H. Tessmer, P.I. Glicofridis, R.C. Ashoori, L.S. Levitov, and M.R. Melloch, 
Nature {\bf 392}, 51 (1998);
G. Finkelstein, P.I. Glicofridis, R.C. Ashoori, and M. Shayegan, 
Science {\bf 289}, 90 (2000);
G. Ben-Shach, A. Haim, I. Appelbaum, Y. Oreg, A. Yacoby, and B.I. Halperin,
Phys. Rev. B {\bf 91}, 045403 (2015).





\bibitem{paper_prb}
M. Pletyukhov, D.M. Kennes, J. Klinovaja, D. Loss, and H. Schoeller, arXiv:1911.06886, 
submitted to Phys. Rev. B. 



\bibitem{kallin_halperin_prb_84}
C. Kallin and B.I. Halperin, Phys. Rev. B {\bf 29}, 2175 (1984).



\bibitem{AAH}
Y. Lahini, R. Pugatch, F. Pozzi, M. Sorel, R. Morandotti, N. Davidson, and Y. Silberberg, 
Phys. Rev. Lett. {\bf 103}, 013901 (2009); 
M. Schreiber, S. S. Hodgman, P. Bordia, H. P. L{\"u}schen, M. H. Fischer, R. Vosk, E. Altman, 
U. Schneider, and I. Bloch, Science  {\bf 349}, 842 (2015);
S. Ganeshan, K. Sun, and S. Das Sarma, 
Phys. Rev. Lett. {\bf 110}, 180403 (2013).




\bibitem{SM}
See Supplemental Material, where the parameters used in Fig.~\ref{fig:band_structure} are listed.




\bibitem{dana_jpc_85}
I. Dana, Y. Avron, and J. Zak, J. Phys. C {\bf 18}, L679 (1985).

\bibitem{kohmoto_prb_89_jpsj_92}
M. Kohmoto, Phys. Rev. B {\bf 39}, 11943 (1989);
{\it ibid.}, J. Phys. Soc. Jpn. {\bf 61}, 2645 (1992).

\bibitem{hatsugai_prb_93}
Y. Hatsugai, Phys. Rev. B {\bf 48}, 11851 (1993).



























\bibitem{multi-channel}
N. M\"uller, D.M. Kennes, M. Pletyukhov, J. Klinovaja, D. Loss, and H. Schoeller, in preparation.

\bibitem{lin_etal_preprint}
Y.-T. Lin, D.M. Kennes, V. Meden, and H. Schoeller, in preparation.

\bibitem{gangadharaiah_etal_prl_12}
S. Gangadharaiah, L. Trifunovic, and D. Loss, Phys. Rev. Lett. {\bf 108}, 136803 (2012).



\bibitem{weyl}
D.M. Kennes, M. Pletyukhov, K. Piasotski, J. Klinovaja, D. Loss, and H. Schoeller, in preparation.







\end{thebibliography}
\end{document}